# Foundations for reconstructing early microbial life

Betül Kaçar

Department of Bacteriology, University of Wisconsin – Madison, Madison WI, USA

bkacar@wisc.edu

**ABSTRACT**

For more than 3.5 billion years, life experienced dramatic environmental extremes on Earth. These include shifts from oxygen-less to over-oxygenated atmospheres and cycling between hothouse conditions and global glaciations. Meanwhile, an ecological revolution took place. The planet evolved from one dominated by microbial life to one containing the plants and animals that are most familiar today. The activities of many key cellular inventions evolved early in the history of life, collectively defining the nature of our biosphere and underpinning human survival. There is a critical need for a new disciplinary synthesis to reveal how microbes and their molecular systems survived ever changing global conditions over deep time. This review critically examines our current understanding of early microbial life and describes the foundations of an emerging area in microbiology and evolutionary synthetic biology to reconstruct the earliest microbial innovations.

## 1. INTRODUCTION

Contemporary microbiology is organized around a complex trove of phenomena: the vast array of molecular machines that power cellular life and the interactions between cohorts of cells that couple to and resonate with their environments (and each other) in unique ways. Despite the current eukaryote-skewed distribution of terrestrial biomass (7), the history of life on our planet is overwhelmingly one of bacteria and archaea, extending back almost 4 billion years (127; 161).

There are two main repositories of information to explore the temporal dimension of microbial history: the geological record of very old rocks, and the bioinformatic content and comparative physiologies of extant species (68; 126; 132; 133; 153). Historical questions have fallen largely to geochemists and paleontologists who focus on the expansive Precambrian stretch of time prior to eukaryote-dominated ecosystems, ~1 billion years ago (127). This division of labor was borne out of a great disparity between specialties required to study these data repositories. The ability to locate, date, analyze and make sense of remnants of once living microbes embedded on old rocks involves tools that are substantially different from the knowledge and techniques required to make sense of the intricacies of cellular function and evolution. The disparity has not been a particularly troublesome issue because researchers in both fields had their respective hands full with pressing problems and an ever-broadening array of tools and materials at their disposal. At the same time, molecular biology and microbiology has played a limited role in reconstructing the planetary history of life. As a field it has traditionally focused on organizing and applying knowledge describing the *extant* expression of microbial interactions, with an emphasis on cellular and subcellular mechanisms.

This *status quo* may now be shifting. Accessibility to and discovery of rocks of great age relevant to characterizing geochemical information about the Earth's earliest microbiota is reaching a plateau as the surface of our planet has mostly been mapped in its entirety. At the same time, bioinformatics and synthetic biology tools that could help to explore the vast temporal dimension of microbial evolution are much more powerful, widely available, and easier to use than they once were. These tools can be applied to many different cellular and metabolic processes in ways that provide new insights into earliest biological development that are not possible through conventional paleontology. Importantly, ancient biochemical solutions that evolved in response to massive environmental and ecological changes may open new biotechnology applications of microbiology to characterizing and potentially mitigating current upheavals in our planetary ecosystem. In light of these changes, it is time to consider how disciplinary boundaries may be crossed to deepen the study of early microbial evolution (219).

This review is intended to serve as a primer for biologists interested in applying their expertise to the study the temporal development of microbial life on our planet. It summarizes our current understanding of the history and evolution of microbial life with emphasis on metabolism; critically reviews the recent advancements towards rebuilding the earliest steps of life in the laboratory. This article may also provide a guide for geologists and geochemists curious about how to develop and apply evolutionary bioinformatics and experimental tools and techniques to questions of our planet's deep past in new ways. It begins with an overview of what makes questions of early microbial evolution challenging, and outlines recent approaches taken to piece together our microbial history across a variety of different disciplines. It finally synthesizes a vision forward as to how the study of the distant past can inform the most urgent challenge of characterizing microbial responses to environmental change.

## 2. A Tale of Two Records
### *2.1. Precambrian Geology*

Our planet has existed for over 4.5 billion years. Rocks have been structurally and chemically altered by biological activity and prevailing environmental conditions at Earth's surface for almost 4 billion years of this time. The geological record provides direct evidence of past environmental conditions and indicates changing degrees and types of biological activity. Though geology focuses on the study of rocks, the advent of nuclear chemistry through the mid-20th century opened entirely new means of conducting geological and paleontological research (190). Nuclear chemistry established methods of determining the absolute ages of most rocks that are accessible at the Earth's surface, for measuring and tracking atomic compositions and isotopic abundances, and for linking fine details of geochemistry to changes in major planetary reservoirs such as the ocean and the atmosphere (189). The rock record is by no means complete and is in many respects biased towards preserving evidence of the recent past in comparison to the deep past. This is because the longer a rock has existed, the likelier it is that it will be weathered into dust or chemically altered due to exposure to the surface environment (18). Consequently, the rocks that survive from Earth's earliest history reflect only a diminishing, surviving fraction of the total volume of rocks from that time period. Any view of Earth's early history is further biased toward organisms and environments that tend to be found in large sedimentary basins formed at the edges of continents (105). By contrast, exposed surface environments and open ocean volumes are poorly represented. Each alteration of an original rock, or of the fossil remains that it might host, introduces an additional scientific caveat that must be studied and understood before any conclusions may be drawn. The complexity of these caveats mainly accounts for why Precambrian paleontology and paleobiology has traditionally been conducted within the realm of geosciences (92).

The rock record, being skewed toward the more recent past, also skews toward a corresponding biological record of eukaryotes because their remains and activities generate more obvious paleontological and chemical traces (124). Nearly all paleontological studies are organized and conducted at the organismal level and are therefore additionally skewed to this level of the biological hierarchy. Much of the business of paleontological work is focused upon interpreting fossil indicators in the context of organismal taxonomic resolution.

The changes in evolutionary biology wrought by the integration of molecular genetics throughout the mid-20th century had a profound impact on paleontology (91; 196-198; 206). By reconciling evolutionary statistics and cladistics with the record of taxonomic, ecological and environmental changes, paleontologists were able to develop and test new hypotheses about the ways in which fundamental evolutionary phenomena such as speciation and adaptation took place. Statistical studies of evolution as recorded in the paleontological record are pinned to metrics associated with attributes that can be traced through time and space, or in relative abundance within an ecosystem (133). As a result, this subject was developed according to the study of terrestrial and shallow-marine, eukaryotic (mostly plant and animal) life using clades possessing numerous, varied and *temporally trackable* morphological characters (199).

The incorporation of molecular genetics, population biology, and evolutionary biology approaches into the study of Precambrian biota is a challenging hurdle. The overwhelming bulk of Earth's biological history, in contrast to its latest chapters, was created by microbial organisms possessing relatively simple external morphologies. Microbial life, lacking most or all such observable characteristics provides a poor basis for historical reconstruction based on paleontological data. Under the rare circumstances where microbial fossil preservation is possible, only the barest outlines of cellular morphology remain with no evidence of internal organelles. There are almost no taxonomically definable characteristics of bacteria or archaea (or even protistan-grade eukaryotes) that derive from morphology (127). This simplicity means that key elements of Earth's biological past are poorly recorded or missing altogether from the geological record.

Despite these limitations, the late 20th century did bring major, fundamental advancements in our reconstruction of planetary microbiology: the oldest conclusive evidence for life was established (~3.4 Ga stromatolites, or layered, sedimentary features formed by microbial communities) (103); the oldest evidence of taxon-specific metabolic activity (GOE, ~2.5 Ga) (69) and corresponding oldest taxon-specific body fossils (cyanobacteria 1.9 Ga) (123), and the oldest eukaryotes (1.5 Ga) (2) and oldest multicellular red and green algae (1 Ga) (33; 82) reflecting the timing of key microbiologically significant endosymbiotic events (see Figure 1). This coarse framework served as a starting point for more elaborate biogeochemical characterization into the 21st century, and it remains largely in place today. It also provides key temporal anchor points for the extension of extant bioinformatics into the deep past.

### 2.2. Modern Bioinformatics Extended to Ancient Microbiology

Where detailed refinement of the geological and paleontological record through the 20th century largely dovetailed with the advent of nuclear chemistry, the bioinformatic record expanded with the advent of computation-assisted chemical and sequence analysis. The chemical structure of genes was established in 1940, the first artificial gene was synthesized in 1979 (10), and the first full gene sequence of a microbial organism was obtained in 1995 (71; 95). The first major application of diverse gene sequences to the systematic assessment of organismal diversity and relative evolutionary relation (phylogenetics) revealed the distinct domains of life (237). Phylogenetics highlighted major deficiencies in cladistic (trait-based) categorization of microbial organisms. The transition to the 21st century saw the construction of the first major databases and the development of sophisticated pipelines and structural inference models that mapped genetic sequences to phenotypical biomolecular functions. Metagenomics has provided almost countless additional sequences that help to fill out this puzzle and extend the understanding of metabolic capacities in every direction (95; 211), including functional integration of biogeochemistry and microbes both at the molecular and global scale (155; 211).

The development of phylogenetics was a watershed moment in the study of microbial evolution that helped to establish the foundations of microbial history, such as the mitochondrial and plastid endosymbiotic events (179). Statistical comparison also enabled the first ways in which bioinformatics, through ancestral sequence reconstruction, was extended into a temporal dimension in ways that complement the limitations of an organism-centric fossil record (53; 60; 165; 210). The excellent overview "What genomes have to say about the evolution of the Earth" (25) summarizes the efforts up to 2012.

## 2.3. *Past is the prelude: The concept of uniformitarianism and ancestral gene reconstruction*

The default context in which almost any interpretation of an ancient phenomenon is made is uniformitarianism: the assumption that the rates, mechanisms, pathways, processes, etc. that are observed in extant life or in the geosphere today can serve as a guide for their ancient counterparts (134). This concept is often colloquially summarized as 'The present is the key to understanding the past.' Uniformitarianism has worked reasonably well for reconstructing the history of the last ~500 million years, for which there are many comparable features of the biosphere and geosphere to be found around us today for comparison. But it becomes increasingly tenuous in application the further back one goes in time. As James Valentine put it in an essay of that name, the 'Infusion of Biology into Paleontological Research' has only just begun to grapple with the microbiotic and biochemical interactivity afforded by the modern synthesis as extended to the Earth's deep microbial history (225). This infusion resulted in the establishment of the broader discipline of 'paleobiology', which is distinguished from paleontology by explicitly including how life and environment have changed throughout Earth's history that go beyond the information preserved by body and trace fossils (see Figure 2).

The ways in which biology interacts with and is shaped by the environment extend far below the level of organisms to realms of biochemistry and cellular physiology. These are levels where the rock record falls short but which the modern tools of bioinformatics can open for study. There are myriad means of coupling biomolecules and environmental substrates for even the 'simplest' microorganisms (27; 174). This makes extant databases of bioinformation pivotal to reconstructing our planet's microbiological history.

Coupling and extrapolating backwards from extant life to make sense of profound geochemical level changes of the deep past is what is conveyed by the term *molecular paleobiology*, and it is at the forefront of current methods in evolution, microbiology, paleobiology and the geosciences. For example, the richness of modern molecular biology and bioinformatics tools offers an alternative pathway for investigating the extent to which uniformitarian assumptions may reasonably serve as a null hypothesis. Reconstruction of ancestral proteins can be used to go further by experimentally testing hypotheses about Earth-life history, namely by constraining past protein functionality and linking functional models to independent insights from geological record (12; 78) (see Figure 1). This computational technique, referred to as ancestral sequence reconstruction or paleogenetics, leverages extant genomic information and statistical models of protein evolution to infer the amino acid sequence composition of an ancestral protein. The gene encoding the ancestral protein can then be synthesized and engineered within a laboratory microbial organism for expression and experimental characterization (see Figure 3) (113; 114). Such experimental efforts form at a confluence of Precambrian geochemistry, genetic information, protein modeling, synthetic biology and evolutionary models. In this way, bioinformatic databases and experimental tools open the ability to create and test new and increasingly sophisticated hypotheses that connect the biotic and surface planetary environments in profound ways.

When the research area to computationally infer Precambrian proteins first opened, questions mainly involved discussing which of relatively few available well-sequenced protein families were better or poorly suited to Precambrian-age questions, and sorting out the extent of confounding or limiting aspects of applying those few systems (25; 60). Research questions involving ancient proteins and deep microbiological history have since become correspondingly much more diverse (12; 25; 78; 85; 90; 139). Geochemical characterization methods have also become more sensitive and nuanced, opening new questions and the development of new ties between geochemistry and biochemistry. Experimental studies involving ancestral proteins have begun to lead to the formulation of foundational questions such as:

- Under what range of environmental conditions are ancient proteins functional? Do these align with or depart from what has been inferred for the surface environment from the geological record? How have changes in abiotic/biotic nutrient cycling influenced the evolution of early enzymes and metabolisms?

- What were the precipitating factors in the origins of novel protein functionality (particularly those functions that had major global consequences, e.g., oxygenic photosynthesis, biological nitrogen fixation, or the rise of multicellularity/eukaryotes)? Under what selective environmental conditions did these proteins emerge and persist?

- What are the ranges of biosignatures generated by ancient proteins? Are they consistent with that produced by modern proteins and how do they shift in response to inferred, ancient environmental conditions?

- What are examples of alternative evolutionary pathways for ancient proteins that are not sampled or expressed in bioinformatic databases? Can they be revealed by synthetic biology and/or experimental evolution studies?

### 2.4. *Current Work: A Protein-Level View of Paleobiology*

Microbial metabolism affects the planet's geochemistry in myriad ways, most notably by altering the redox state of metals, imparting subtle changes in the proportions of stable isotopes, and by synthesizing and altering the composition of organic compounds hosted within rocks (5). Correspondingly, there is a wide array of bioinformatic study designs using protein and nucleotide sequences to reconstruct life's history (54; 127). (See Table 1 for a summary of recent studies that have applied ancestral sequence reconstruction to investigate the evolution of geochemically significant enzymes and metabolisms). At the same time, the development of new and more sophisticated analytical geochemical techniques has provided more nuanced means of linking the sparse rock record to the inferred biochemical history of ancient cells (84; 135; 178; 216).

The experimental work following the phylogenetic reconstruction of the ancestral sequences will depend on the protein of study and the research question. For example, many proteins can be recombinantly expressed in common model organisms that may not be relevant to deep time studies and thus may fall outside the desired scope to assess the functionality of the protein relative to geochemical change over time. In some cases, if tools are available, it may be preferable to directly engineer an ancient gene or gene clusters into the microbial genome. Such genetic level engineering would introduce physiological change which, if carefully designed, can be interpreted to result directly from the different functionality of the expressed ancient protein (**Figure 3**). A recent methods protocol by Garcia, Fer and others provides a thorough step-by-step guideline for utilizing ancestral sequence reconstruction in paleobiological studies, and summarizes potential pitfall and mitigation strategies of the ancestral sequence reconstruction approach (76).

Regardless of field of study or methods employed, over the last few years these works have expanded our understanding of sub-organismal (i.e., metabolic, or biosynthetic) historical microbiology. They are gathered here for comparison to review how research groups focus on ancient proteins and metabolic pathways that have changed over time. Changes in these pathways provide profound insights into the deep history of microbial life to complement the limited fossil remains that characterize conventional Precambrian paleontology.

### 3. Carbon Metabolism
#### 3.1. *Carbon Isotope Fractionation*

The centrality of organic carbon to biochemical structures has placed carbon isotope fractionation at the center of interpretations of biological primary productivity over the entire history of life on our planet (73). Carbon, essential for life, is fixed from inorganic compounds (e.g., $CO_2$) into organic molecules by a handful of known metabolic pathways upon which the productivity of the biosphere depends (119). From each of these metabolic pathways, carbon isotope patterns form due to the preferential uptake of the lighter carbon isotope, $^{12}C$, versus the heavier stable isotope $^{13}C$, at the active sites of enzymes. These patterns can vary based on the catalytic properties of the enzymes involved, the physiological characteristics of the host organism, and environmental conditions (75). A sample of organic matter can be tested for biological origins by comparing its $^{13}C/^{12}C$ ratio to a reference point for abiotic carbon, which is typically the carbon found in carbonate rocks from the same sample. A consistent carbon isotope fractionation pattern of about -20 to -30 per mil (the standard representation of the ratio of $^{13}C$ to $^{12}C$, or $\delta^{13}C$, indicating 'parts per thousand' of a fractional difference compared to an abiotic standard) can be traced back almost as far back as there are geological materials to analyze, almost 3.8 billion years (75). Analysis of ancient $^{13}C$ isotope signatures can 1) provide evidence for the past presence of (carbon-fixing) biological activity and 2) constrain candidate carbon fixation pathways responsible for generating these signatures. Carbon isotope signatures are generally regarded as some of the most robust geochemical indicators of biological activity available (75; 100), but this generality comes at the expense of specific insight into the taxonomy of the organisms that produce such fractionation values (44).

No rocks older than ~4 billion years in age remain today. The only geochemical traces from the prior period of Earth history are recycled, individual mineral grains hosted in younger rocks (43; 226) that in certain cases have small inclusions of carbon-bearing matter (11; 98; 154). The state of the art for interpreting these earliest possible geochemical indicators of life is attempting to reconcile these incredibly sparse carbon isotope data with other isotopic or geochemical indicators consistent with biological activity or, more generally, habitability (11; 44; 56; 57; 149; 184).

In recent years, as the microbial and biochemical (i.e., suborganismal) picture of carbon fixation has become more detailed, studies have turned to focus on specific enzymes, physiological structures and pathways thought to be most responsible for imparting carbon isotope fractionation at sub-organismal levels (23). Central to many of these studies is the enzyme Ribulose 1,5-bisphosphate (RuBP) carboxylase/oxygenase, EC 4.1.1.39 (RuBisCO) of the Calvin-Benson-Bassham carbon fixation pathway. RuBisCO catalyzes the uptake of inorganic $CO_2$ from the environment and facilitates $CO_2$ reduction and incorporation into organic biomass (6). Although there are at least six known distinct pathways of carbon fixation, each driven by many different metabolic enzymes, RuBisCO is currently the only known carbon metabolic enzyme that generates the characteristic fractionation value observed between abiogenic carbonate and biological organic carbon for the last few billion years (75) importantly though methanotrophic metabolism, and both biotic and abiotic methane production can also produce $^{13}C/^{12}C$ ratios outside of the range of carbon isotopic signatures found in the geologic record (see Figure 2) (170; 184; 228). As a result, much of the effort put toward reconciling the Precambrian carbon isotope fractionation evolution via molecular biology and phylogenetics has started with the RuBisCO ancestral sequence reconstruction and phylogenetic studies (116; 120; 205). These efforts have been foundational to exploring the evolutionary history of RuBisCO, namely by establishing the extent to which we may relate extant functionalities to earliest properties (120; 231). Recent related efforts have looked beyond RuBisCO to investigate the impacts of other cellular physiological, anatomical or enzymatic effects on isotope fractionation, such as carbon concentration mechanisms in cyanobacteria (108); more such efforts are expected to follow.

### 3.2. Methanogenesis

Like atmospheric carbon dioxide, methane is also a key geochemical source of carbon for biomass production (47). Both gases contribute to the greenhouse effect (the retention of heat from sunlight that leads to the warming of our planet's surface), but of the two gases methane is the more potent greenhouse gas. In the deep past, our sun generated about 10-20% less sunlight than today, which most planetary models indicate should lead to a plunging of temperatures to well below the freezing point of water (118). There is no geologic evidence of widespread freezing of the planet's surface in the deep past, so the implication is that the atmosphere had enough of a greenhouse effect that it prevented widespread glaciation (or possibly that there is some other geological or biological factor maintaining a habitable planet that escapes our understanding). Carbon dioxide alone is not thought to have been present in sufficient amounts to maintain temperatures above freezing, so it has been inferred that powerful greenhouse gases such as methane made up for the difference. For these reasons, reconstructing or constraining the origins of metabolic methanogenesis is critical for understanding the habitability and stability of our planet's surface (130).

The claim for the oldest, direct geochemical evidence of life arises from extremely low values of $^{13}C$ isotope fractionation measured in hydrothermally produced, methane-bearing fluid inclusions dated around 3.47 Ga (223). There have been a range of studies attempting to estimate the timing of methanogenesis emergence in a broader context of retracing the origins of microbial emergence and divergence and using estimated rates of genetic change. Depending on the organismal gene datasets, specific molecular evolutionary models and age constraints employed, estimates for earliest methanogenesis range from 3.05–4.49 Ga (8), 3.46–3.49 Ga (140), 2.97–3.33 Ga (203) and slightly younger ranges of around 2.75–3.17 Ga for hyperthermophilic methanogenesis in an ancestral Euryarcheota (14). A recent study takes a novel approach to this question by leveraging horizontal gene transfer (HGT) events between a clade with no clear geochemical record (Archaeal methanogens) and one with a more tractable Precambrian history (Cyanobacteria). Molecular clock analyses calibrated by using HGT events as constraints indicate that methanogens diverged within Euryarchaeota no later than 3.51 billion years ago, with methanogenesis itself probably evolving earlier (238). Taken together, most of the models indicate support for scenarios wherein microbial methane production was important in maintaining temperatures consistent with liquid water on the early Earth.

### 3.3. Protometabolic Inferences

One of the strengths of taking a bioinformatic approach to carbon metabolism is that it enables the formulation and testing of new hypotheses that are not explicitly tied to evidence from the geochemical record. Basic questions about the earliest history of life abound, including debates about the likeliest phylogenetic placement of the root of the bacterial lineage within the tree of life (48) and age correlations between the divergence of microbial domains and the occupation of marine and terrestrial niches (9; 107). There have also been fundamental efforts to characterize the full scope of different pathways of carbon uptake and fixation used by all extant organisms, and then to ask what the most parsimonious order of origination of these pathways might have been. For example, Braakman and Smith (28) reconstructed a phylometabolic tree with a high degree of parsimony that traces the evolution of carbon-fixation pathways down to its root, which consisted of a reductive citric acid cycle with the Wood-Ljungdal pathway forming a simple network. This linked network lacks the selective optimization of modern fixation pathways, but its redundancy leads to a more robust topology, making it more plausible than any modern pathway as a primitive universal ancestral form of biological carbon fixation. They also argued that elaboration of this basic root appeared to result from selection for energy conservation and oxidative toxicity. In a similar vein, other researchers have looked at the full scope of metabolism and investigated whether there is evidence of a central 'protometabolic' network that could reflect properties of a last universal common ancestor (LUCA). Most such efforts have focused on assessing a wide range of sequenced and annotated microbial genomes and analyzing a core subset of metabolic reactions (49; 83; 239). Nevertheless, some efforts have gone further in attempting to peer into environmental adaptation (37) and horizontal gene transfer events prior to the establishment of LUCA (72), to assess the likeliest external conditions that hosted LUCA-like organisms (234; 235) or even to design carbon-catalyzing artificial chemistries inspired by inferred protometabolic pathways (51).

4. **Phosphorus Metabolism**

Though it composes only a few percent of dry body mass, phosphorus is absolutely essential to the emergence and early evolution of life as we know it (64; 67; 152; 221). It is thought to have been a significant limiting nutrient for most, if not all, of our planet's microbiological history. Phosphorus atoms touch on all three autocatalytic realms of cellular life (74): in metabolism it serves as the primary means of storing and transducing energy harvested from diverse free energy gradients; in information transfer it forms the structural backbone of ribonucleic and deoxyribonucleic polymers; and in cell membranes it forms a key structural component of amphiphilic lipids (94).

Despite its centrality to biochemical functionality, it is not clear that phosphorus could have been found in abundant quantities at the time of life's origins and early evolution. The most common (oxidized) forms of phosphate are noted for low aqueous solubility (164) and there are no known sources of geochemically widespread or long-lifetime sources of reduced phosphorus (221). In light of these limitations, it has been hypothesized that more abundant and soluble sulfur compounds could have played primordial roles as cofactor, electron donor and electron acceptor in place of problematic phosphorus (63; 99). As just one example of how this progression may have occurred, acetyl phosphate is hypothesized to be a bridge between predecessor thioester and phosphate metabolisms (236). Simpler thioesters, with equivalent functional chemistry to the modern cofactor acetyl CoA, would have formed the basis for a transition between prebiotic chemistry and extant metabolic networks in cells (169).

Computational analyses of metabolic networks have been leveraged to develop and test hypotheses about sulfur's primordial role as an energy transducer that preceded or complemented phosphorus. A bioinformatic mapping of cellular metabolism can be stripped of reactions that require phosphorus. The remaining network forms a coherent core of reactions that can exist independently of phosphate activity but are nevertheless able to support a host of major metabolic functions using thioester-linked metabolites and iron- and sulfur-containing enzymatic cofactors (88; 89; 142). These analyses indicate that a primordial thioester-based metabolism could plausibly have preceded phosphorylated biomolecules (99).

Beyond its roles in enabling (or perhaps constraining) the origins of life, recent efforts to model Precambrian geochemical phosphorus availability have highlighted some surprising implications for phosphorus in controlling the development of our planetary environment. When phosphorus acts as such a limiting substrate for global biological productivity, small shifts in the fate of biomass can lead to significant changes in fluxes through carbon and oxygen reservoirs (81; 180; 222). The result is that in the deep past, phosphorus could exert outsized effects on surface conditions (namely, oxygen partial pressure and surface temperature regulation) through feedback mechanisms that today are muted by massive terrestrial biomass reservoirs and oxygenated atmospheres and oceans (122). One such study highlighted how anoxic oceans inhibited the capacity for remineralization of phosphorus obtained from biomass, thereby hampering primary productivity and suppressing oxygenation of our planet by hundreds of millions of years (122). Another possible mechanism with a similar overall effect would have stemmed from enhanced phosphorus scavenging in anoxic, iron-rich oceans and a nutrient-based bistability in atmospheric oxygen levels. This would have resulted in elemental stoichiometries in primary producers that diverged strongly from the Redfield ratio (the 106:16:1 atomic ratio of carbon:nitrogen:phosphorous in marine phytoplanktonic biomass) and a protracted, low-oxygen world (173). An elaboration on these models would add sulfate as a switch for phosphorus regeneration from biomass trapped in anoxic sediments. Low-sulfate conditions in the early and middle Precambrian would lead to muted phosphorus regeneration, while high-sulfate conditions characteristic of some parts of the late Precambrian would drive increased rates of anoxic phosphorous regeneration that could be part of a shift in the overall surface system to a new stability point (66; 121). The clear implication, regardless of the mechanism studied, is that microbial (re)processing of phosphorus played an outsized role in modulating the surface environment for billions of years.

5. **Nitrogen Metabolism**

Breaking the triple bond of dinitrogen gas found ubiquitously in our atmosphere is an extremely energy-intensive process, requiring 251 kJ/mol for activation (150) or, in biotic terms, 16-40 ATP molecules for enzymatic reduction (96). The ability to break nitrogen gas and then incorporate it into bioavailable ammonia, a process known as nitrogen fixation, only evolved once in the known history of this metabolism and is carried out by the nitrogenase enzyme (26; 86; 176) that is strictly dependent on ATP (97). In other words, all enzymes that can carry out the nitrogen fixation function can be mapped on a single phylogenetic network. Many nitrogenase enzymes use a molybdenum-containing iron-sulfur cofactor, which is somewhat unusual given molybdenum's scarcity in terrestrial environments and in the bulk marine environment prior to Earth's atmospheric oxygenation ~2.5 Ga (80). Prior to the emergence of nitrogenase, it seems likely that global bioproductivity would have been limited (perhaps severely so) by the amount of reduced nitrogen generated by non-enzymatic phenomena such as UV photolysis, lightning discharges, or background particle radiation sources such as galactic cosmic radiation, solar flares or radioactive mineral seams (151).

A comparison between phosphorus and nitrogen is insightful. Though both are considered likely candidates as limiting substrates throughout the Precambrian history of life, the singular ability of nitrogenase to reduce nitrogen opens a possibility to reconcile nitrogenase phylogenetics with an independent record of nitrogen biogeochemical cycling (163; 176; 214). This has led to parallel debates in both geochemistry and in molecular evolution circles about the prospects of timing nitrogenase origins and subsequent diversification (163; 176).

The use of nitrogen isotopes to reconstruct a geochemical history of nitrogen cycling is a relatively new technique. Nitrogen isotope ratios from marine and river sedimentary rocks deposited between 3.2 and 2.75 billion years ago are relatively close to values generated by modern nitrogen-fixing prokaryotes and are specifically consistent with values generated by molybdenum-dependent nitrogen fixation (213). These ancient sedimentary materials are less depleted in the light $^{14}$N isotope than fixed nitrogen generated by nitrogenase variants with cofactors containing vanadium or additional iron in place of molybdenum. These data cannot readily be explained by abiotic processes and therefore suggest earliest biological nitrogen fixation was performed by molybdenum-based nitrogenase (213). These proposed Archean nitrogen fixation signatures contrast with the wider range of isotopic values observed from the Proterozoic, coincident with rising atmospheric oxygen and concomitant increased cycling of more oxidized nitrogen species (212; 214).

Against this geochemical context, the phylogenetic history of nitrogenase is a subject of ongoing work and debate. The indications of a 3.2-billion-year-old molybdenum-based nitrogenase is at odds with scarce molybdenum geochemical availability in ancient, Archean oceans. This has led to nitrogenase phylogenetic and computational protein structural analyses designed to test hypotheses stating that variants only utilizing iron, a highly abundant metal in Archean oceans, may have preceded those utilizing molybdenum (26; 80; 145). For example, a competing view based on phylogenetic analyses suggests that molybdenum-based nitrogenase ancestral variants had characteristics of maturases that can scaffold molybdenum cofactors and preceded iron- or vanadium-based variants despite molybdenum's lower marine availability early in Earth history (79; 80). An additional interesting observation that stems from the molybdenum-first model is that nitrogenase variants with iron- and vanadium-based cofactors also possess a key protein subunit (the G-subunit) that is an example of an "orphan" gene – a sequence for which there is no obvious counterpart or predecessor anywhere else in biology (50). The orphan G-subunit seem to play a critical role in accommodating the cofactor functions of these other metals (46; 50; 187), and its emergence may be coincident following the Great Oxidation Event approximately 2.5 billion years ago (163). Another model proposes that acquisition of nitrogenase by planktonic, unicellular cyanobacterial lineages was delayed until the very end of the Precambrian (about 800-600 million years ago), coincident with 'Snowball Earth' events, periods when much of the planet's surface had frozen over (182). The timing of this model is mostly consistent with a view that nitrogenase diversification among key, planktonic primary producers would have been stalled by a lack of molybdenum and vanadium in the marine environment for the preceding 3+ billion years, and that the corresponding ripple effects of increased primary productivity afforded by lifting of a nitrogen bottleneck are implicated with leading to the Snowball Earth events themselves.

6.  **Photoreceptor and Photosystem Evolution**

The sun's contribution to our planet's surface energy budget is vast (~300 Watts per square meter). This energy helps to maintain a habitable surface temperature and powers many basic chemical reactions that convert abiotic elemental sources into biomass. The directionality of the energy input can also be leveraged by microbial organisms to increase their chances of survival in different ecological and physiological contexts. In these different ways, the development of photochemical biological structures has become indispensable to the sustained presence of life on our planet (146). The phylogenetic relationships of photochemical systems to one another, and their coupling to tractable changes in geochemical reservoirs, are at the forefront of reconstructing our planet's microbiological history.

*6.1 Anoxygenic Photosynthesis*

Phototrophism (the ability to convert the free energy of photons into metabolically-available chemical energy) represents the greatest contributor to ecosystem energy budgets (131). Without phototrophism, microbes can certainly make use of electron donors and acceptors to make a living in isolated niches. For example, microorganisms recovered from isolated aquifers can survive for prolonged periods using the redox power provided by subsurface ionizing radiation (160). But phototrophy removes free energy as a limiting constraint on the size of the biosphere while enabling access to new ecological niches that can facilitate organismal variability. The origin of phototrophy was thus considered as one the most significant events in life's history (148). The available evidence indicates that phototrophism has existed for most (if not all) of the apparent history of life on our planet (103). Phototrophism is made possible by enzymatic photochemical reaction centers (RCs) that generate and channel excited electrons, which are passed through a chain of transport enzymes to a final electron acceptor. The electron potential generated in this way produces reduced compounds that can be used to power an array of cellular biosynthetic reactions.

Reconstructing the detailed history of phototrophism has proven to be extremely challenging, but the broad outlines of this history are beginning to take shape. Photosynthetic RCs are categorized into two classes—Type I and Type II (40). From biophysical and sequence analysis of RCs, it has been understood that the proteins within these two classes are structurally and functionally similar, each likely descending from a single common ancestor (17). All known RCs have a dimeric core of proteins. In most cases, these are heterodimers that consist of two subunits (16; 192). The similarity of the two halves of the heterodimeric complexes in both sequence and structure, indicates that both classes of photosynthetic RCs derive from a single, primordial common ancestor originating amongst anoxygenic photosynthetic lineages, which then later diversified via gene duplication and divergence events (17). Recent studies of the web of cofactors spread across bacterial taxa indicate that the primordial phototrophic ancestor of all extant bacterial phototrophs must have had both Type I and Type II reaction centers (38).

Only the barest outlines of phototrophism's origins are recorded in the geochemical record. There is a well-documented presence of shallow marine stromatolitic structures (interpreted to evidence photosynthetic microbial activity) that predate the rise of atmospheric oxygen (or any indication of oxygenation), extending back over the last ~3.5 billion years of time (229). There are also abundant deposits of the magnesium carbonate mineral, dolomite, for much of the anoxic and suboxic history of our planet, with textures that indicate that this mineral precipitated at the same time as the formation of stromatolitic features (24). Until recently, though, there were no obvious, environmentally widespread ways to precipitate dolomite at volumes commensurate with its presence in ancient deposits. However, recent microbial studies have indicated that unique biogeochemical circumstances created by anoxygenic photosynthetic microbial ecosystems can facilitate the precipitation of dolomite (52). There was a dramatic shift in shallow marine geochemistry and atmospheric composition at ~2.5 billion years ago indicating a massive increase in the partial pressure of oxygen (the Great Oxidation Event or GOE), and a significant and ongoing debate about indications of oxygen in the few hundreds of millions of years that preceded the GOE (4; 207).

This geochemical context makes a case that anoxygenic photosynthesis preceded oxygenic photosynthesis by as much as a billion years, and that the coupling of two different photosystems to enable oxygen production represented a substantial increase in microbial photosystem complexity and capability (70; 104; 183). This basic understanding remains mostly unchanged, but a confluence of different pieces of data in recent years has opened new views into the pre-oxygenic history of phototrophism and indicate a more complicated picture. Anoxygenic photosynthesis is restricted to members of the bacterial domain, but the major phototrophic lineages are not closely related to one another. This evolutionary distance obscures reconstructive efforts with long divergence times between clades and low sequence similarity across key enzyme factors. One approach compared phylogenetic relationships based on a photosynthesis tree (constructed from sequences mainly centered on type-II photosynthetic reaction center RC enzymes such as PufHLM and BchXYZ), which were compared with those of 16S rRNA sequences (109). Many aspects of the topologies of these trees were congruent. The phylogenetic relations demonstrated that bacteriochlorophyll biosynthesis had evolved in ancestors of phototrophic green bacteria earlier as compared to phototrophic purple bacteria and that multiple events independently formed different lineages of aerobic phototrophic bacteria (109). In a related effort, a novel method was developed to track the relative placement of horizontal gene transfer events across photosynthetic lineages, and stem green non-sulfur bacteria were likely the oldest phototrophic lineages (138).

New evidence regarding anoxygenic phototrophism continues to accumulate that could inform such efforts. Recently discovered putative phototrophs of the uncultivated candidate phylum "WPS-2" encode a distinct lineage of anoxygenic Type II reaction centers and unusual characteristics such as the capacity for aerobic respiration, reflecting a divergent evolutionary history and likely novel ecological roles (232). Additionally, a phylogenetic analysis of the duplication event that led to the evolution of the core antenna subunits CP43 and CP47 can be compared with more recent events such as the duplication of Cyanobacteria-specific FtsH metalloprotease subunits and the radiation leading to closely related clades, including those containing anoxygenic phototrophs (22). Ancestral state reconstruction in combination with comparative structural biology of photosystem subunits provides additional evidence supporting the premise that water oxidation had originated before the ancestral core duplications, with subsequent specialization of the enzymes making up the photosystems leading to oxyphototrophy (39; 41; 159). This possibility challenges current thoughts on the evolution of photosynthesis by indicating that the history of pre-oxygenic photosynthesis is far more complicated than has generally been appreciated, and may further be amenable to more detailed reconstructive analysis or laboratory experimentation (158).

Combining these observations with the relative development of anoxygenic and oxygenic photosynthetic molecular systems, and the clear phylogenetic and structural relationships between them, it is possible that the hundreds of millions of years of anoxygenic phototrophic evolutionary activity that predate the rise of oxygen may be resolvable at the sequence level despite the paucity of the rock record from this time.

*6.2. Timing Oxygenic Photosynthesis, Crown Group Cyanobacteria and the Rise of Atmospheric Oxygen*

The emergence of oxygenic photosynthesis is implicated with the significant rise of atmospheric oxygen approximately 2.4 billion years ago (the aforementioned GOE). Separately, there is a broadly consistent history of distinct cyanobacterial body fossils going back about 2 billion years ago (one of vanishingly few bacterial lineages with tractable morphological characteristics). Cyanobacterial fossils are complemented by a paleontological history of anatomically sophisticated (chloroplast-hosting) red- and green-algal descendants that first appear approximately 1 billion years ago (33; 82), which trace a direct path to the emergence and radiation of even more sophisticated green land plants around 450 million years ago. The inheritance and subsequent radiation of cyanobacteria-derived plastids and free-living organisms can be comparatively studied with a correspondingly complex and relatively well-constrained phylogenetic history of eukaryotes in ways that are impossible for most other microbial taxa.

A reasonably parsimonious assumption would be that the emergence of oxygenic photosynthesis and the emergence of crown group Cyanobacteria temporally coincided with the first dramatic rise of atmospheric oxygen (the GOE) (55). By contrast, there are many different clades with variants of anoxygenic photosynthesis, each utilizing cofactors and RCs with common predecessors found in oxygenic photosystems (181). As with the reconstructed history of anoxygenic photosynthesis above, recently acquired data and more detailed phylogenetic analyses indicate a more complicated picture.

Cyanobacteria branched off long ago from other bacterial lineages (102) and until recently, no closely related taxa were known. Molecular surveys of animalian gut and subsurface aquifer habitats led to the discovery of uncultured lineages that could be phylogenetically classified as members of the Cyanobacteria, which were subsequently categorized into a small number of neighboring phyla (59; 209). Metabolic analyses suggest that their common ancestors were non-photosynthetic, anaerobic, motile, and obligately fermentative (59). A wider sampling of genomes of additional major sibling groups of Cyanobacteria indicate that while clades such as Melainabacteria have diverse energy metabolisms (including fermentation and aerobic or anaerobic respiration), $H_2$ metabolism is a common feature of the modern groups that cannot respire aerobically, suggesting that hydrogenases may have played key roles in ancestral metabolisms (144).

These new data have helped to distinguish between the details of the mechanistic history of metabolic oxyphototrophy from the overlay of host organism phylogenetics. A phylogenomic approach derived from a well-resolved 'core' cyanobacterial tree, combined with ancestral state reconstruction, indicates that early cyanobacteria were probably unicellular and restricted to freshwater habitats with small cell diameters until at least 2.4 Ga (15). In this view, the GOE may be associated less strongly with the emergence of oxyphototrophy than with the expansion of the metabolic capability to marginal and marine habitats and the ability to form thick, stabilizing microbial mats; both factors which significantly increased biosphere-level flux through the oxyphototrophy pathway (185; 186) with concomitant effects on atmospheric reservoirs. Recent studies have built on these insights. An assessment of horizontal gene transfer events that affected the distribution of reaction centers across phototrophic taxa improved the precision of phototroph divergence date estimates and was used to test among different evolutionary models that connect anoxygenic and oxygenic phototrophy. This approach yielded age ranges for crown Cyanobacteria from 2.2-2.7 Ga and stem Cyanobacteria diverging ~2.8 Ga, several hundred million years prior to the GOE (138). A separate approach using cross-calibrated Bayesian relaxed molecular clock analyses show that crown group oxyphotobacteria evolved only 2.0 billion years ago, with an estimated divergence between oxyphotobacteria and Melainabacteria ca. 2.5–2.6 Ga, which would likely mark an upper limit for the origin of oxygenic photosynthesis (204).

The rise of atmospheric oxygen simultaneously opened new pathways for oxygen mitigation and consumption whose histories can also be studied using bioinformatic approaches. Oxyphototrophic ancestors generated oxygen from water, which may have alleviated a key bottleneck for access to an electron transfer substrate. However, many key enzymes found in anoxic metabolism were still extremely sensitive to oxygen. Rather than focus on the history of its photosystems, another approach lies in studying changes in superoxide dismutase enzymes (SODs) capable of removing superoxide free radicals (62; 171). Bayesian molecular clocks predict that stem group Cyanobacteria may have arisen as early as 3300–3600 million years ago, and that this much deeper metabolic history could have been indirectly shaped by the availability of key metallic cofactors such as nickel, manganese or iron leveraged to neutralize reactive oxygen species (22). Net oxygen accumulation may also be implicated with increased burial or removal of organic matter from ecosystem processing (thereby reducing or significantly slowing the 'back reaction' of biomass with oxygen to produce $CO_2$). There are an array of geological and biological candidate mechanisms promoting organic matter burial with the capacity to affect global reservoirs (13). As just one example, patterns of diversification within flavin-dependent Baeyer–Villiger monooxygenases have been implicated with multiple pulses of atmospheric oxygenation, suggesting that global oxygenation processes may be more fully characterized as systemic autocatalytic transitions rather than simply overproduction through oxyphototrophic pathways (201). Phylogenetic analyses that focus just upon oxygen-utilizing and oxygen-producing enzymes intriguingly indicate that oxygen metabolism may have significantly predated the GOE, with diverse oxygenases and oxidoreductases emerging around 3.1 billion years ago (111).

Clearly, elucidating the historical interconnections between oxygen production and its attenuation in the cellular environment represents a noteworthy area of promise, particularly when these interconnections may be coupled to trends in inorganic substrate availability (143; 176) .

*6.3. Photosensitive Pigments*

Alongside phototrophic molecular systems that directly harvest the free energy of incident photons for metabolism, there are separate histories of much simpler photosensitive molecules and accessory pigments. The most notable among these is rhodopsin, which has been coopted by biota in all domains (Archaea, Bacteria, and Eukarya) to serve an array of functional roles such as ion pumps, enzyme catalysis and photosensing (65; 93). The functional versatility of rhodopsin across diverse taxa (some of which have well-documented fossil histories) has conferred new ways of constraining the history of a molecule that, on its own, leaves no other direct or indirect geochemical traces.

Rhodopsin is a class of proteins whose common features are a seven-transmembrane alpha-helix apoprotein and a cofactor of retinal (32; 137; 168; 240). Rhodopsin distribution across such diverse taxa has been attributed to ancient origins, possibly predating the divergence of the three domains of life (177), or to high rates of microbial gene transfer (202). The sensitivity of key residues to the modulation of specific peak spectral absorbance is a particularly useful feature of rhodopsins that can be extrapolated to the study of its phylogeny (117) and microbial ecology (101).

Ancestral sequence reconstruction has been used to study the history of rhodopsins in a variety of different contexts, typically regarding its host organisms and their ecological niches. Its oldest variants likely acted as light-driven proton pumps (200). Consistent with an inferred ancestral function as a proton pump, the common ancestors of microbial rhodopsins were spectrally tuned toward the absorption of green light, which would have enabled its hosts to occupy depths in a water column or biofilm where UV wavelengths were attenuated (195). Diversification of rhodopsin functions and peak absorption frequencies likely coincide with an expansion of surface ecological niches induced by the accumulation of atmospheric oxygen (195). The phylogeny and inferred functions of bacterial rhodopsins as biochemical entities can be studied despite the lack of any corresponding fossil record of its hosts, potentially covering spans of billions of years.

Eukaryotic rhodopsins and visual opsins used by eukaryotes with well-documented fossil histories, in contrast to bacterial rhodopsins, allows for very specific inferences about the behaviors of their eukaryotic hosts (45; 61; 172; 208; 227). Maximum likelihood phylogenetic ancestral reconstruction methods have been used to infer the amino acid sequence of a putative ancestral archosaur visual pigment that was possessed by a diapsid reptile living about 200 million years ago. Artificial genes expressed in mammalian cell lines yielded photoactive pigments with maximum absorption values at a wavelength of about 508 nm, which is slightly redshifted relative to that of extant vertebrate pigments (45). In another study, resurrected ancestral cetacean (whale) rhodopsins indicate that the ancestor of modern cetaceans dived to deep marine depths, exhibiting both a deep-sea spectral shift and accelerated retinal kinetics consistent with rapid adjustment to dimming light (61).

7. **Lipid Synthesis and Metabolism**

Studies of the history of lipid synthesis represent some of the newest conceptual ties between respective advances in both microbiology and geochemistry. All cells are bound by phospholipid membranes to prevent diffusion of enzymatic catalysts and genetic material, and to mediate cellular interactions with the surrounding environment such as the acquisition of metabolic substrates, expulsion of waste products and the spatial coordination of cellular attachment and taxis (132). Bacterial and eukaryotic phospholipids employ the same membrane biochemistry based on glycerol-3-phosphate, though almost all eukaryotes produce distinct tetracyclic sterols that regulate membrane physiology (233). By contrast, a wide range of bacteria synthesize the structurally related pentacyclic hopanepolyols (31). Archaeal phospholipids are based on glycerol-1-phosphate linked to chains of isoprenoids (35). Fossilized lipids preserved in sedimentary rocks can offer insights into ecological trends across high-level clades when unique lipid synthesis pathways can be mapped onto distinct extant organisms (3; 21; 31; 132; 217; 218; 241). An exhaustive summary of lipid paleobiological 'biomarkers' is outside the scope of this review, but recent studies have shown that bioinformation-based reconstructive approaches may offer potentially new insights into the interpretation of these biomolecules, specifically with respect to tracking the earliest origins of eukaryotes (141; 218).

Eukaryotes emerged following the GOE, with the earliest complex fossils consistent with basic eukaryotic attributes (large size, complex morphology and complex cell wall ultrastructure) first showing up around 1.6 billion years ago (2). Although these morphological features are known only in eukaryotes, it is unclear whether they indicate stem- or crown-group eukaryotes in the absence of preserved subcellular organelles. The last common ancestor of all eukaryotes was an oxygen-respiring organism and possessed a mitochondrion (125), and molecular clock estimates based on broad phylogenomic sampling across major eukaryotic lineages have indicated that it emerged between 1.8-1.6 billion years ago (162). This implies a substantial lag of several hundred million years between the rise of oxygen associated with the GOE and the earliest suspected eukaryotic fossils, with correspondingly poor resolution of the intervening evolutionary steps and environmental circumstances of their emergence (128). The relative proportion of eukaryotes to bacterial and archaeal clades in these ecosystems is uncertain, since fossil counts may be subject to any number of observational or preservational biases. With these unknowns and uncertainties, eukaryote lipid diversity has come to be regarded as an independent, complementary tool for tracking eukaryote emergence and early evolution.

A systematic assessment of Precambrian rocks has been underway for at least the last 30 years in an attempt to reconcile fossil occurrence patterns with biomarker composition. Recovered lipids suggest yet another intriguing lacuna of almost a billion years between the earliest eukaryotic microfossils and the first occurrences of (eukaryotic) 4-desmethyl steranes in marine rocks from around 800 million years ago (29; 110; 241). This period of time broadly coincides with an exceptionally long period of depressed eukaryote fossil diversification and ecological turnover (34; 167). It is difficult to make sense of these data, as prolonged evolutionary stases on 100+ million-year time scales challenges most reasonable expectations of evolutionary mode and tempo inferred from observations of extant protistan-grade organisms. One possibility is that the data generally reflect Earth system history, and that ~800 million years ago marked a sharp ecological transition from bacteria-dominated microbial mat ecosystems to eukaryote-dominated communities shaped by algae and heterotrophic feeding strategies. In this case, we might expect that early eukaryotes differed in some substantial way from their extant counterparts. Another possibility is that low abundances of steranes before this time indicate widespread preservational bias, with most eukaryotic steroids being destroyed or consumed within bacteria-dominated microbial mats (166).

Recent data lend some credence to the first possibility that early eukaryotes indeed differed substantially from extant taxa. Geochemically preserved lipids covering the period from about 1600-800 million years ago lack diagnostically eukaryotic steranes, but they do include an array of 'protosteroid' derived compounds that are not employed by extant taxa. The compositions of these diverse protosterols are consistent with early intermediates of modern sterol biosynthesis pathways (19; 20), which would seem to place substantial limitations on characterizing the biochemical (and biochemistry-derived morphological, anatomical and evolutionary) capabilities of microbiota from this time period using only body fossil synapomorphies.

The discovery of divergent lipid protosterols opens new possibilities for investigating the opaque history of eukaryotes through a combination of bioinformatic and biosynthetic techniques. Desmond and Gribaldo (58) conducted a pioneering study of the taxonomic distribution and phylogeny of sterol pathway enzymes across taxa, finding evidence that bacteria acquired analogs of 'modern' sterol synthesis capabilities from eukaryotes via HGT. Gold et al. (87) used molecular clock techniques to focus more specifically on timing the origins of sterol biosynthesis by studying the key enzymes squalene monooxygenase and oxidosqualene cyclase, both of which rely upon molecular oxygen as a substrate to produce protosterol precursors. They found that the maximum marginal probability for the divergence time of this system is broadly coincident with the GOE at around 2.31 Gyr ago. They also found that these two key enzymes recapitulate organismal trees fairly well and were likely present in the last eukaryotic common ancestor. These two attributes would indicate a reasonable possibility for reconstructing the circumstances of ancestral intermediate protosterol synthesis using modern host organisms, perhaps up to and including engineering organisms with a capacity for synthesizing non-canonical lipids. Studies of lipid synthesis may be complemented by corresponding studies of ancestral lipid metabolism (e.g., (175)). The prospects for piecing together the vast amounts of early eukaryotic evolutionary history using body fossils are declining, but those for utilizing a combination of phylogenomics, gene engineering and geochemical kerogen recovery are only getting started.

### 8. The Past is the Future: Applying Early Microbiology to Pressing Problems

The picture emerging over the last few years of work is one in which the deep history of taxonomic relationships, gene transfer events, endosymbiotic acquisitions, biochemical structures and inferred functions of metabolic processes can be leveraged to explore the subcellular and biochemical macroevolutionary history of our planetary biology. The emphasis in these approaches is notably shifted away from organisms and populations and data available through bioinformatic and geological repositories are both valuable and yet sparse for their own reasons. A biosynthesis-based approach, based on molecular evolution and interpreted in a microbiological context, is well-suited to making sense of fragmentary geochemical data that exist from a time of microbiotic hegemony that is defined less by anatomical complexity and more by biosynthetic and gene-regulatory innovation. The field is shifting. Current work is focused on developing an array of bioinformatic techniques spanning from ancestral network reconstructions to machine-learning-guided protein design, and upon selecting combinations of gene engineering systems and microbial hosts that are best suited to addressing key questions for which data are partially or entirely occluded.

This is not to say that efforts along these lines will remain focused on understanding the past. Every organism alive today is a product of ~4 billion years of life's evolutionary past on our planet and the history of our life is one of absolute survival in the face of temperature extremes, catastrophic impacts, massive upheavals such as increases or decreases in available substrates and nutrients, or dramatic shifts between marine and terrestrial niches. Organisms have had billions of years and countless generations of tinkering to sort out how best to make a biological living. They have sampled sequence and environmental combinations that vastly outstrip our capacity of laboratory experimentation and field observation. Planetary microbiology, in practice, represents the development of shortcuts to likely-functional variants that are no longer widely expressed in modern organisms. What has changed today is our ability to track key aspects of these adaptations, which are indelibly imprinted in the histories of biosynthetic systems that persisted through these challenges.

The bioinformatic resource of life's remarkable evolutionary history can be applied in ways that address critical societal problems by integrating microbiology with our knowledge of Earth and its history. Precambrian protein reconstructions provide one way to bridge fundamental questions of microbiology with generic, planetary-scale processes in ways that go beyond the historical study of Earth and may include a broader array of biotechnological applications. Future studies may be designed explicitly to uncover novel uses or inter-molecular relationships that are not found in extant organisms but may have been found in the deep past or will be a possible outcome of evolution in the future. Advances in (and to some extent, standardization of) synthetic biology techniques will help advance the field to one in which ancient protein reconstruction studies can be parallelizable in a high-throughput manner, including but not limited to rebuilding ancient systems in cell-free compartments for desired biotechnological applications. These tools will give researchers the ability to efficiently test a broad range of hypotheses and to target multiple molecular systems of interest.

Life's diversity is a testament to its ability to solve problems of survival and biosynthesis in unique ways, and life's history is an unparalleled repository of molecular innovations in the face of environmental extremes. The greatest value in understanding the past may lie in unraveling biologically vetted possibilities for dealing with uncertainties of our own future.

**Acknowledgments**

I thank my colleagues and friends in molecular evolution, genetics, synthetic biology, astronomy, planetary sciences and geology communities for shaping my thinking about life, time and planets over the years. I am deeply grateful for all the members of my young laboratory who courageously took over bold problems and old proteins, as well as the members of the NASA MUSE ICAR team and Katrina Forest for the critical reading of the manuscript. I thank the NASA Science Mission Directorate and the NASA Astrobiology Program, National Science Foundation, John Templeton Foundation, Hypothesis Fund, Gates Ventures, and Human Frontiers in Science Program for generously supporting this work. Finally, I thank the editors of the Annual Reviews in Microbiology for the invitation.

**Table 1.** Select recent studies that have employed ancestral sequence reconstruction to investigate the evolution of ancient microbial metabolisms and biogeochemistry

| Target protein | Target function | Reference[a] | Laboratory reconstruction? | Genetic engineering? |
|---|---|---|---|---|
| Methyl-coenzyme M reductase | Methane metabolism | Adam et al. 2022 (1) | No | No |
| | | McKay et al. 2019 (147) | No | No |
| Nitrogenase | Nitrogen fixation | Harris et al. 2024 (97) | Yes | Yes |
| | | Cuevas-Zuviría et al. 2024 (50) | Yes | Yes |
| | | Garcia et al. 2023 (77) | Yes | Yes |
| Oxidosqualene cyclase | Steroid biosynthesis | Hoshino and Gaucher 2021 (106) | Yes | No |
| Photosystem II | Oxygenic photosynthesis | Oliver et al. 2021 (159) | No | No |
| Rhodopsin | Ion transport | Sephus et al. 2022 (195) | No | No |
| RuBisCO | Carbon fixation | Kedzior et al. 2022 (120) | Yes | Yes |
| | | Schulz et al. 2022 (193) | Yes | No |
| | | Wang et al. 2023 (230) | Yes | Yes |
| | | Amritkar et al. 2024 (242) | No | No |
| Superoxide dismutase | Oxygen protection | Sendra et al. 2023 (194) | Yes | No |
| | | Valenti et al. 2022 (224) | Yes | No |

[a]Studies that used other methods of reconstructing properties of ancient metabolisms, including ancestral state reconstruction and consensus reconstruction have not been included.

**Figures and captions**

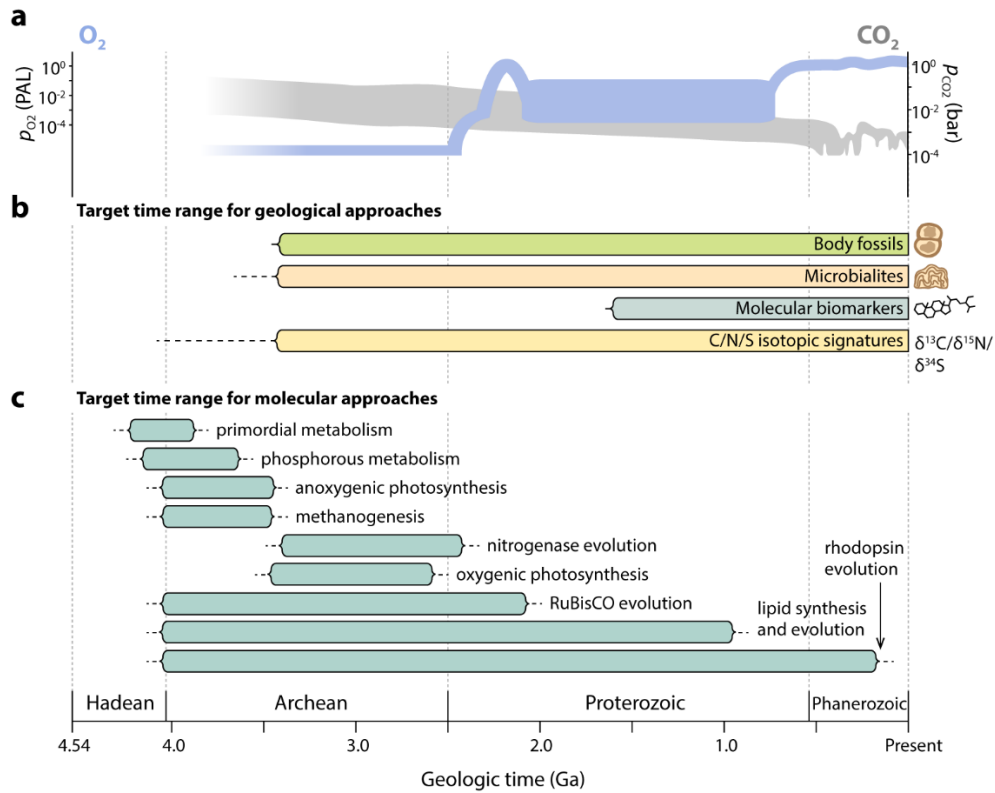

**Figure 1.** Geological and molecular records of life and environment on Earth. *(a)* Reconstruction of past atmospheric oxygen ($O_2$) and carbon dioxide ($CO_2$) levels, adapted from Lyons et al. (136) and Catling and Zahnle (42), respectively. *(b)* Periods in Earth history for which paleobiological reconstruction is possible via geological study of preserved biosignatures (body fossils, microbialites, molecular biomarkers, and isotopic signatures). Time ranges for biosignatures are synthesized from the following references (dotted lines indicate where biogenicity is not fully established): body fossils (112; 157; 188; 191; 215), microbialites (103; 112; 156; 191), molecular biomarkers (30; 218), carbon/nitrogen/sulfur isotopic signatures (11; 36; 149; 184; 214; 220). *(c)* Major periods of metabolic evolution and diversification provide tractable targets for molecular/microbiological methods of paleobiological reconstruction. Bars represent approximate time ranges associated with the evolutionary events and critical open questions discussed in the main text (these are not solely intended to depict ages of emergence for the listed metabolisms).

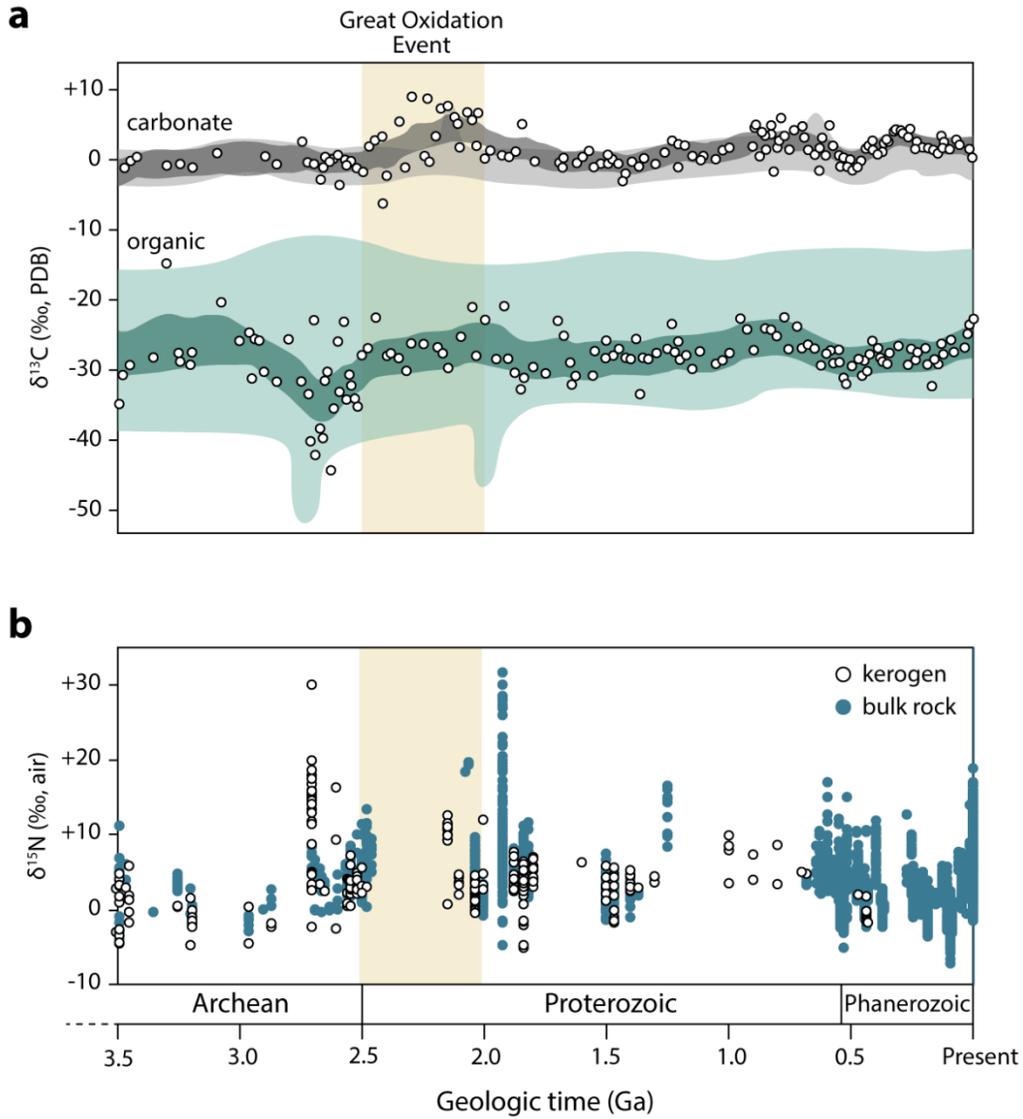

**Figure 2**. Stable isotopic records of *(a)* carbon (δ¹³C) and *(b)* nitrogen (δ¹⁵N) from preserved geological materials. *(a)* Nitrogen isotopic data points representing individual measurements are compiled from Stüeken et al. (214) and references therein. *(b)* Carbon isotopic data points are compiled from Krissansen-Totton et al. (129) and references therein, which represent binned (10 Mya) values from individual measurements. Dark grey and dark green fields, modified from Krissansen-Totton et al. (129), represent 95% confidence intervals from a Kalman smoother fit to both carbonate and organic measurement values. Light grey and light green fields represent full ranges of carbon isotopic measurements, modified from Schidlowski (184).

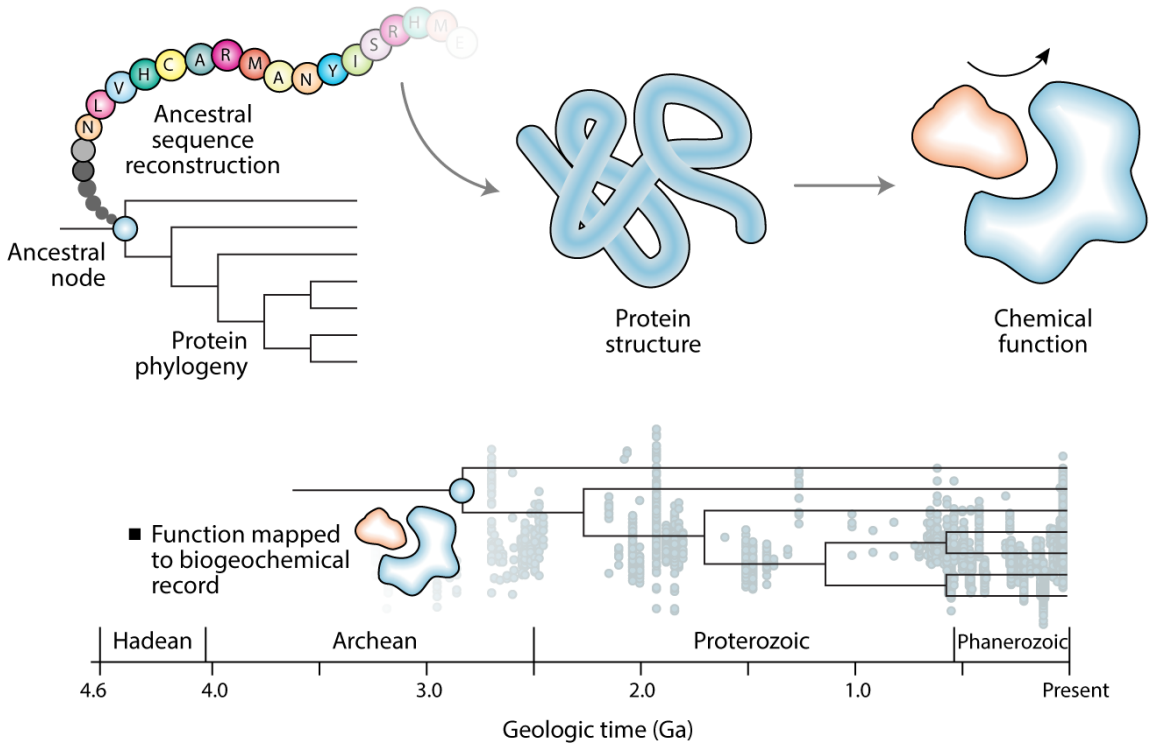

**Figure 3.** Historical molecular sequences, which can be reconstructed via genomic and phylogenetic tools, can yield experimentally verified insights into ancient protein structure and chemical function. Function can then be mapped to independent biogeochemical records of life (78; 115).

40